\documentclass[aps,prl,superscriptaddress,showpacs,floatfix,twocolumn]{revtex4}
\usepackage{amsmath}
  \usepackage[dvips]{graphicx}

\def\Journal#1#2#3#4{{#1}{\bf #2}, #3 (#4)}

\def\NIMA{{Nucl. Instrum. Methods}~{\bf A}}
\def\NPA{{Nucl. Phys.}~{\bf A}}

\def\PLB{{Phys. Lett.}~{\bf B}}
\def\PLC{Phys. Repts.\ }
\def\PRL{Phys. Rev. Lett.\ }
\def\PRD{{Phys. Rev.}~{\bf D}}
\def\PRC{{Phys. Rev.}~{\bf C}}
\def\ZPC{{Z. Phys.}~{\bf C}}

\begin{document}

\title{Measurement of Non-Random Event-by-Event Fluctuations of Average
Transverse Momentum in $\sqrt{s_{NN}}=200$ GeV Au+Au and p+p Collisions}

\newcommand{\abilene}{Abilene Christian University, Abilene, TX 79699, USA}
\newcommand{\acadsin}{Institute of Physics, Academia Sinica, Taipei 11529, Taiwan}
\newcommand{\banaras}{Department of Physics, Banaras Hindu University, Varanasi 221005, India}
\newcommand{\barc}{Bhabha Atomic Research Centre, Bombay 400 085, India}
\newcommand{\bnl}{Brookhaven National Laboratory, Upton, NY 11973-5000, USA}
\newcommand{\caucr}{University of California - Riverside, Riverside, CA 92521, USA}
\newcommand{\ciae}{China Institute of Atomic Energy (CIAE), Beijing, People's Republic of China}
\newcommand{\cns}{Center for Nuclear Study, Graduate School of Science, University of Tokyo, 7-3-1 Hongo, Bunkyo, Tokyo 113-0033, Japan}
\newcommand{\columbia}{Columbia University, New York, NY 10027 and Nevis Laboratories, Irvington, NY 10533, USA}
\newcommand{\dapnia}{Dapnia, CEA Saclay, F-91191, Gif-sur-Yvette, France}
\newcommand{\debrecen}{Debrecen University, H-4010 Debrecen, Egyetem t{\'e}r 1, Hungary}
\newcommand{\fsu}{Florida State University, Tallahassee, FL 32306, USA}
\newcommand{\gsu}{Georgia State University, Atlanta, GA 30303, USA}
\newcommand{\hiroshima}{Hiroshima University, Kagamiyama, Higashi-Hiroshima 739-8526, Japan}
\newcommand{\ihepprot}{Institute for High Energy Physics (IHEP), Protvino, Russia}
\newcommand{\isu}{Iowa State University, Ames, IA 50011, USA}
\newcommand{\jinrdubna}{Joint Institute for Nuclear Research, 141980 Dubna, Moscow Region, Russia}
\newcommand{\kaeri}{KAERI, Cyclotron Application Laboratory, Seoul, South Korea}
\newcommand{\kangnung}{Kangnung National University, Kangnung 210-702, South Korea}
\newcommand{\kek}{KEK, High Energy Accelerator Research Organization, Tsukuba-shi, Ibaraki-ken 305-0801, Japan}
\newcommand{\kfki}{KFKI Research Institute for Particle and Nuclear Physics (RMKI), H-1525 Budapest 114, POBox 49, Hungary}
\newcommand{\korea}{Korea University, Seoul, 136-701, Korea}
\newcommand{\kurchatov}{Russian Research Center ``Kurchatov Institute", Moscow, Russia}
\newcommand{\kyoto}{Kyoto University, Kyoto 606, Japan}
\newcommand{\labllr}{Laboratoire Leprince-Ringuet, Ecole Polytechnique, CNRS-IN2P3, Route de Saclay, F-91128, Palaiseau, France}
\newcommand{\lawllnl}{Lawrence Livermore National Laboratory, Livermore, CA 94550, USA}
\newcommand{\losalamos}{Los Alamos National Laboratory, Los Alamos, NM 87545, USA}
\newcommand{\lpc}{LPC, Universit{\'e} Blaise Pascal, CNRS-IN2P3, Clermont-Fd, 63177 Aubiere Cedex, France}
\newcommand{\lund}{Department of Physics, Lund University, Box 118, SE-221 00 Lund, Sweden}
\newcommand{\muenster}{Institut f\"ur Kernphysik, University of Muenster, D-48149 Muenster, Germany}
\newcommand{\myongji}{Myongji University, Yongin, Kyonggido 449-728, Korea}
\newcommand{\nagasaki}{Nagasaki Institute of Applied Science, Nagasaki-shi, Nagasaki 851-0193, Japan}
\newcommand{\newmex}{University of New Mexico, Albuquerque, NM, USA}
\newcommand{\nmsu}{New Mexico State University, Las Cruces, NM 88003, USA}
\newcommand{\ornl}{Oak Ridge National Laboratory, Oak Ridge, TN 37831, USA}
\newcommand{\orsay}{IPN-Orsay, Universite Paris Sud, CNRS-IN2P3, BP1, F-91406, Orsay, France}
\newcommand{\pnpi}{PNPI, Petersburg Nuclear Physics Institute, Gatchina, Russia}
\newcommand{\riken}{RIKEN (The Institute of Physical and Chemical Research), Wako, Saitama 351-0198, JAPAN}
\newcommand{\rkrbrc}{RIKEN BNL Research Center, Brookhaven National Laboratory, Upton, NY 11973-5000, USA}
\newcommand{\saispbstu}{St. Petersburg State Technical University, St. Petersburg, Russia}
\newcommand{\saopaulo}{Universidade de S{\~a}o Paulo, Instituto de F\'{\i}sica, Caixa Postal 66318, S{\~a}o Paulo CEP05315-970, Brazil}
\newcommand{\seoulnat}{System Electronics Laboratory, Seoul National University, Seoul, South Korea}
\newcommand{\stonybrkc}{Chemistry Department, Stony Brook University, SUNY, Stony Brook, NY 11794-3400, USA}
\newcommand{\stonycrkp}{Department of Physics and Astronomy, Stony Brook University, SUNY, Stony Brook, NY 11794, USA}
\newcommand{\subatech}{SUBATECH (Ecole des Mines de Nantes, CNRS-IN2P3, Universit{\'e} de Nantes) BP 20722 - 44307, Nantes, France}
\newcommand{\tenn}{University of Tennessee, Knoxville, TN 37996, USA}
\newcommand{\titech}{Department of Physics, Tokyo Institute of Technology, Tokyo, 152-8551, Japan}
\newcommand{\tsukuba}{Institute of Physics, University of Tsukuba, Tsukuba, Ibaraki 305, Japan}
\newcommand{\vandy}{Vanderbilt University, Nashville, TN 37235, USA}
\newcommand{\waseda}{Waseda University, Advanced Research Institute for Science and Engineering, 17 Kikui-cho, Shinjuku-ku, Tokyo 162-0044, Japan}
\newcommand{\weizmann}{Weizmann Institute, Rehovot 76100, Israel}
\newcommand{\yonsei}{Yonsei University, IPAP, Seoul 120-749, Korea}
\affiliation{\abilene}
\affiliation{\acadsin}
\affiliation{\banaras}
\affiliation{\barc}
\affiliation{\bnl}
\affiliation{\caucr}
\affiliation{\ciae}
\affiliation{\cns}
\affiliation{\columbia}
\affiliation{\dapnia}
\affiliation{\debrecen}
\affiliation{\fsu}
\affiliation{\gsu}
\affiliation{\hiroshima}
\affiliation{\ihepprot}
\affiliation{\isu}
\affiliation{\jinrdubna}
\affiliation{\kaeri}
\affiliation{\kangnung}
\affiliation{\kek}
\affiliation{\kfki}
\affiliation{\korea}
\affiliation{\kurchatov}
\affiliation{\kyoto}
\affiliation{\labllr}
\affiliation{\lawllnl}
\affiliation{\losalamos}
\affiliation{\lpc}
\affiliation{\lund}
\affiliation{\muenster}
\affiliation{\myongji}
\affiliation{\nagasaki}
\affiliation{\newmex}
\affiliation{\nmsu}
\affiliation{\ornl}
\affiliation{\orsay}
\affiliation{\pnpi}
\affiliation{\riken}
\affiliation{\rkrbrc}
\affiliation{\saispbstu}
\affiliation{\saopaulo}
\affiliation{\seoulnat}
\affiliation{\stonybrkc}
\affiliation{\stonycrkp}
\affiliation{\subatech}
\affiliation{\tenn}
\affiliation{\titech}
\affiliation{\tsukuba}
\affiliation{\vandy}
\affiliation{\waseda}
\affiliation{\weizmann}
\affiliation{\yonsei}
\author{S.S.~Adler}	\affiliation{\bnl}
\author{S.~Afanasiev}	\affiliation{\jinrdubna}
\author{C.~Aidala}	\affiliation{\bnl}
\author{N.N.~Ajitanand}	\affiliation{\stonybrkc}
\author{Y.~Akiba}	\affiliation{\kek} \affiliation{\riken}
\author{J.~Alexander}	\affiliation{\stonybrkc}
\author{R.~Amirikas}	\affiliation{\fsu}
\author{L.~Aphecetche}	\affiliation{\subatech}
\author{S.H.~Aronson}	\affiliation{\bnl}
\author{R.~Averbeck}	\affiliation{\stonycrkp}
% ...
\collaboration{PHENIX Collaboration} \noaffiliation

\date{\today}        

\begin{abstract}
Event-by-event fluctuations of the average transverse momentum of
produced particles near
mid-rapidity have been measured by the PHENIX Collaboration in
$\sqrt{s_{NN}}=200$ GeV Au+Au and p+p collisions at the Relativistic Heavy
Ion Collider.  The fluctuations are observed to be in excess of the
expectation for statistically independent particle emission for all
centralities. The excess fluctuations exhibit a dependence on both the
centrality of the collision and on the $p_T$ range over which the average
is calculated.  Both the centrality and $p_T$ dependence can be well
reproduced by a simulation of random particle production with the
addition of contributions from hard scattering processes.

\end{abstract}

\maketitle

%Introduction

The measurement of fluctuations in the event-by-event average transverse
momentum of produced particles in relativistic heavy ion collisions has 
been proposed as a probe of phase instabilities near the QCD phase 
transition~\cite{Heis01,Step98,Step99}, which could result in classes 
of events with different properties, such as the effective temperature 
of the collision. Fluctuation measurements could also provide information 
about the onset of thermalization in the system~\cite{Gav03}. The
resulting phenomena can be observed by measuring deviations of the
event-by-event average $p_T$, referred to here as $M_{p_T}$, of produced
charged particles from the expectation for statistically independent
particle emission~\cite{Stod95,Shur98} after subtracting contributions from
fluctuations arising from physical processes such as elliptic flow and jet
production.

Several $M_{p_T}$ fluctuation measurements have been reported in heavy ion
collisions~\cite{NA49,CERES,PPG005,STAR}, including a study by 
PHENIX~\cite{PPG005} in $\sqrt{s_{NN}}=130$ GeV Au+Au collisions which set 
limits on the magnitude of non-random fluctuations in $M_{p_T}$. Recently, STAR
has reported fluctuations in excess of the random expectation, within the
PHENIX limits, at the same collision energy~\cite{STAR}. 
For the first results from $\sqrt{s_{NN}}=200$ GeV Au+Au and p+p collisions
reported here, upgrades of the PHENIX central arm spectrometers~\cite{PHNIM}
have expanded the azimuthal acceptance from $58.5^\circ$ to 
$180.0^\circ$ within the pseudorapidity range of $|\eta|<0.35$.  Pad chamber 
and calorimeter detectors have also been utilized for improved background 
rejection. As a result, the sensitivity of the PHENIX spectrometer to the 
observation of fluctuations 
in $M_{p_T}$ due to event-by-event fluctuations in the effective temperature 
~\cite{PPG005,Korus01} has improved by greater than a factor of two.

%Data Analysis

Minimum bias events triggered by a coincidence between the Zero Degree
Calorimeters (ZDC) and the Beam-Beam Counters (BBC), with a requirement
that the collision vertex, which is measured with an r.m.s. resolution of less
than 6 mm in central collisions and 8 mm in the most peripheral collisions,
be within 5 cm of the nominal origin, are used in
this analysis.  Event centrality for Au+Au collisions, which is defined 
using correlations in the BBC and ZDC analog response~\cite{PPG002}, 
is divided into several classes, each containing an average of 244,000 
analyzed events.  These classes
are associated to the estimated average number of participants in the
collision, $<N_{part}>$, which is derived using a Glauber model Monte
Carlo calculation with the BBC and ZDC detector response taken into
account~\cite{Glauber}.

Charged particle momenta are reconstructed in the PHENIX central arm
spectrometers with a drift chamber and a radially adjacent pixel pad
chamber. Non-vertex track background rejection is provided by pixel pad
chambers and calorimeters located further outward radially from the
collision vertex~\cite{Mit02}. The momentum resolution is $\frac{\delta
p}{p} \simeq 0.7\% \oplus 1.0\% \times p$ (GeV/c).

$M_{p_T}$ is calculated for each event, which contains a number of 
reconstructed tracks within a specified $p_T$ range, $N_{tracks}$. The $p_T$ 
range is always given a lower bound of 200 MeV/c and a varying upper bound, 
$p_{T}^{max}$, from 500 MeV/c to 2.0 GeV/c. There is a minimum $N_{tracks}$ 
cut of 3 in both Au+Au events (removing 0\%, 4.6\%, and 29\% of events in the 
0-50\%, 50-60\%, and 60-70\% centrality ranges, respectively, when 
$p_{T}^{max}$ = 2.0 GeV/c) and p+p events (removing 59\% of the events).  

There are several measures by which the magnitude of non-random
fluctuations can be quantified, namely $\phi_{p_T}$~\cite{Gaz92,Mro98}, 
$\nu_{dynamic}$~\cite{Pru02}, and $F_{p_T}$~\cite{PPG005}. 
The calculation of $F_{p_T}$ is based
upon the magnitude of the fluctuation, $\omega_{p_T}$, defined as
\begin{equation}
   \omega_{p_T} = \frac{(<M_{p_T}^2> - <M_{p_T}>^2)^{1/2}}{<M_{p_T}>} = \frac{\sigma_{M_{p_T}}}{<M_{p_T}>}.
\end{equation}
$F_{p_T}$ is defined as the fractional deviation of $\omega_{p_T}$ from a baseline estimate defined using mixed events,
\begin{equation}
   F_{p_T} = \frac{(\omega_{(p_T,~data)}-\omega_{(p_T,~mixed)})}{\omega_{(p_T,~mixed)}}.
\end{equation}

Mixed event $M_{p_T}$ distributions are validated by comparisons to a 
calculation of $M_{p_T}$ assuming statistically independent particle 
emission using parameters extracted from the inclusive $p_T$ distributions 
of the data~\cite{Tan01}. For the 0-5\% centrality class, which suffers 
the most from tracking inefficiency, the effects of two-track resolution, and
background contributions, the mixed event $M_{p_T}$ distribution yields 
a value of $F_{p_T}=0.04\%$ with respect to the calculation. The results 
of this comparison are included in the estimates of the systematic errors.
Further details on the mixed event procedure and a discussion of contributions
to the value of $F_{p_T}$ from detector efficiency and resolution effects
can be found in the description of the data analysis of $\sqrt{s_{NN}}=130$ 
GeV Au+Au collisions~\cite{PPG005}.

% Results

Comparisons of the data and mixed event $M_{p_T}$ distributions for the
0-5\% and 30-35\% centrality classes are shown in Fig. 1. Any excess
fluctuations are small and are difficult to distinguish by eye
in a direct overlay of the $M_{p_T}$ distributions. Therefore, the 
comparison is also shown as residuals of the difference
between the data and mixed event distributions in units of standard
deviations of the individual data points. The 
double-peaked shape in the residual distributions is an artifact of the 
fact that the mixed event distributions, which always have a smaller 
standard deviation in $M_{p_T}$ than the data, are normalized to minimize 
the total $\chi^2$ of the residual distribution.

%%%%%%%%%%%%%%%%%%%%%%%%%%%%%%%%%%%%%%%% Figure 1.
\begin{figure}[tb]
\includegraphics[width=1.0\linewidth]{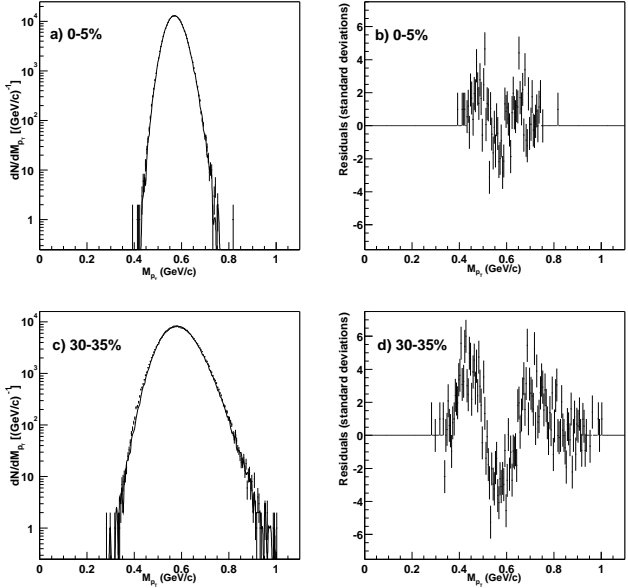}
%\centerline{\epsfig{file=residuals.eps,width=8.5cm}}
\caption[]{\label{fig:1}
Comparisons between the data and mixed event $M_{p_T}$ distributions for 
the representative 0-5\% and 30-35\% centrality 
classes.  Plots a) and c) show direct comparisons of the data (points) and
normalized mixed event (solid line) $M_{p_T}$ distributions.  Plots b) and
d) show the residuals between the data and mixed events in units of
standard deviations of the data points from the mixed event points.}
\end{figure}
%%%%%%%%%%%%%%%%%%%%%%%%%%%%%%%%%%%%%%%%   

Figure 2 shows the magnitude of $F_{p_T}$, expressed in percent, as a
function of centrality for Au+Au collisions with $p_{T}^{max}$ = 2.0 GeV/c. 
The error bars are dominated
by time-dependent systematic effects during the data taking period due to
detector variations, which are minimized using strict time-dependent cuts
on the mean and standard deviations of the inclusive $p_T$ and
$N_{tracks}$ distributions.  
Statistical errors are below $F_{p_T}$ = 0.05\% for all centralities.
The systematic errors are determined by
dividing the entire dataset into ten separate subsets for each centrality
class and extracting the standard deviation of the $F_{p_T}$ values
calculated for each subset.  From Fig. 2, a significant non-random
fluctuation is seen that appears to peak in mid-central collisions.
However, the magnitude of the observed fluctuations are within previously
published limits~\cite{PPG005}.  In addition, the value of $F_{p_T}$ for
the most peripheral Au+Au collisions is consistent with, albeit slightly 
below, the value measured by the same PHENIX apparatus in minimum bias 
$\sqrt{s_{NN}}$ = 200 GeV p+p collisions. If the magnitude of $F_{p_T}$ is 
entirely due to fluctuations in the effective temperature of the 
system~\cite{Korus01}, this measurement corresponds to a fluctuation of 
$\sigma_{T}/<T>$=1.8\% at 0-5\% centrality and 3.7\% at 20-25\% centrality.

%%%%%%%%%%%%%%%%%%%%%%%%%%%%%%%%%%%%%%%% Figure 2. \vspace{1cm}
\begin{figure}[bt]
\includegraphics[width=1.0\linewidth]{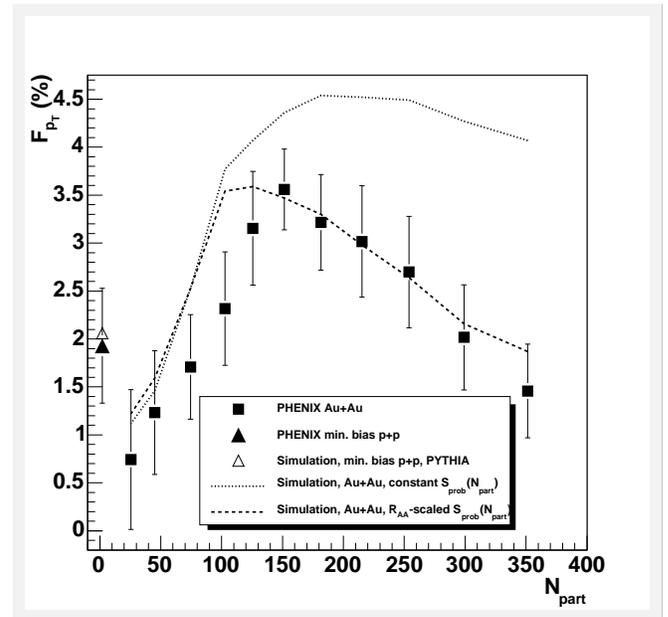}
%\centerline{\epsfig{file=FtCent.eps,width=8.5cm}} 
\caption[]{\label{fig:2}
$F_{p_T}$ (in percent, $0.2$ GeV/c $<p_T<2.0$ GeV/c) as a function 
of centrality, which is expressed in 
terms of the number of participants in the collision, $N_{part}$.  The 
solid squares represent the Au+Au data. The solid triangle represents 
the minimum bias p+p data point. The open triangle is the result from 
an analysis of PYTHIA minimum bias p+p events within the PHENIX acceptance. 
The error bars include statistical and systematic errors and are dominated 
by the latter. The curves are the results of a Monte Carlo simulation with 
hard processes modelled using PYTHIA with a constant (dotted curve) and 
$R_{AA}$-scaled (dashed curve) hard scattering probability factor, and
include the estimated contribution due to elliptic flow.}
\end{figure}
%%%%%%%%%%%%%%%%%%%%%%%%%%%%%%%%%%%%%%%%   

To further understand the source of the non-random fluctuations, $F_{p_T}$
is measured over a varying $p_T$ range for which $M_{p_T}$ is calculated,
$0.2$ GeV/c $<p_T<p_{T}^{max}$.  Figure 3 shows $F_{p_T}$ plotted as a
function of $p_{T}^{max}$ for the 20-25\% centrality class.  A trend of
increasing $F_{p_T}$ for increasing $p_{T}^{max}$ is observed for this and
all other centrality classes.  The majority of the contribution to
$F_{p_T}$ appears to be due to correlations of particles with
$p_T>1.0$ GeV/c, where $F_{p_T}$ increases disproportionately to the
small increase (only 14\%) of $N_{tracks}$ in this region.

%%%%%%%%%%%%%%%%%%%%%%%%%%%%%%%%%%%%%%%% Figure 3.
\begin{figure}[tb]
\includegraphics[width=1.0\linewidth]{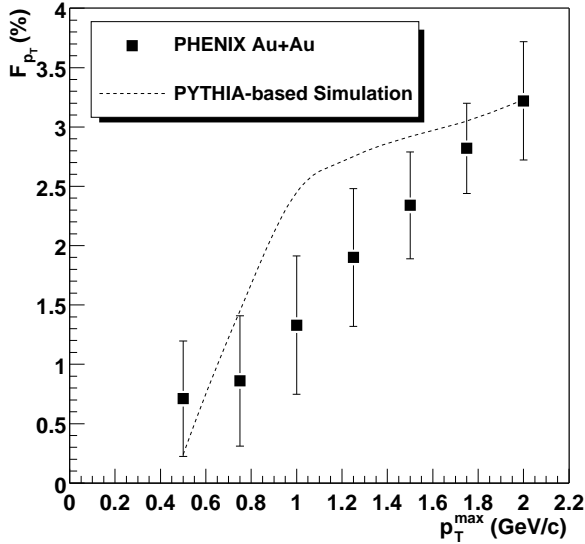}
%\centerline{\epsfig{file=FtPtMax.eps,width=8.5cm}}
\caption[]{\label{fig:3}
$F_{p_T}$ (in percent) of non-random fluctuations as a function of
the $p_T$ range over which $M_{p_T}$ is calculated, $0.2$ GeV/c
$<p_T<p_{T}^{max}$, for the 20-25\% centrality class ($N_{part}$=181.6).
The curve is the result of a
Monte Carlo simulation with hard-scattering processes modelled using PYTHIA
with $S_{prob}(N_{part})$ = 0.075 and $R_{AA}$ = 0.41~\cite{phenixSupp}.
The error bars include statistical and systematic errors and
are dominated by the latter. The contribution of elliptic flow is 
estimated to be negligible at this centrality.} 
\end{figure}
%%%%%%%%%%%%%%%%%%%%%%%%%%%%%%%%%%%%%%%%   

% Contribution of Elliptic Flow

The behavior of $F_{p_T}$ as a function of centrality and $p_T$ is similar
to trends seen in measurements of elliptic flow~\cite{phenixFlow}.  The
contribution of elliptic flow to the magnitude of $F_{p_T}$ is
investigated using a Monte Carlo simulation
whereby events are generated with a Gaussian distribution of $N_{tracks}$ 
particles determined by a fit to the data and a random reaction plane azimuthal 
angle, $\Phi$, between 0 and $2\pi$.  Independent particles within an event
are generated following the inclusive $p_T$ distribution with azimuthal angles,
$\phi$, distributed according to collective elliptic flow described by the function 
$\frac{dN}{d(\phi-\Phi)}=1+2 v_2 cos(2(\phi-\Phi))$.  The values of the
$v_2$ parameter are linearly parameterized as a function of $p_T$ and
centrality using PHENIX measurements of inclusive charged 
hadrons~\cite{phenixFlow}. Only generated particles that lie within the PHENIX
azimuthal acceptance are included in the calculation of $M_{p_T}$.
This simulation estimates that the contribution of elliptic flow to $F_{p_T}$ 
is largely cancelled out by the symmetry of the PHENIX acceptance, and is negligible
for central collisions. The estimated elliptic flow contribution to the value
of $F_{p_T}$ is less than 0.1\% for $N_{part}>150$, increasing to about 0.6\% for 
$N_{part}<100$.  Note that $F_{p_T}$ measured for minimum bias p+p collisions, where 
collective flow is not expected to contribute, is non-zero ($1.9 \pm 0.6\%$), implying 
that a non-flow contribution may also be present in peripheral Au+Au collisions.

% Contribution of Jets

Figure 3 illustrates that a large contribution to the observed non-random
fluctuations is due to the correlation of high $p_T$ particles, such as
might be expected from correlations due to jet production \cite{Liu03}.  
In order to estimate the contribution due to jets, a Monte Carlo simulation 
is again applied.  
Events are generated with a Gaussian distribution of $N_{tracks}$
particles as independent particles that follow an $m_T$-exponential fit
to the inclusive data $p_T$ distribution.  Hard processes are defined to 
occur at a uniform rate per generated particle, $S_{prob}(N_{part})$, for 
each centrality class.  This is the only parameter that is allowed to vary 
in the simulation.  As Au+Au events are being generated, single 
$\sqrt{s_{NN}}=200$ GeV p+p hard-scattering events generated by the 
PYTHIA event generator~\cite{PYTHIA} and filtered by the PHENIX 
acceptance are embedded into the event.  The addition of 
the PYTHIA events affects the mean and standard deviation of the inclusive 
$p_T$ spectra by less than 0.1\%.  The value of $F_{p_T}$ has been extracted 
from 100,000 PYTHIA events for minimum bias p+p collisions, yielding  
$F_{p_T}$=2.06\% within 
the PHENIX acceptance, which is consistent with the measured value of 
$F_{p_T}=1.9 \pm 0.6\%$.

Two scenarios are considered for studies of the centrality-dependence of 
jet contributions to the value of $F_{p_T}$:
1) with $S_{prob}(N_{part})$ set at a constant rate for all centrality classes, 
and 2) with $S_{prob}(N_{part})$ scaled for each centrality class by the PHENIX 
measurement of the suppression of high $p_T$ charged particles, which is characterized 
by the nuclear modification factor, $R_{AA}$, integrated over $p_T>4.5$ 
GeV/c~\cite{phenixSupp}.  The $p_T$ value at which $R_{AA}$ is extracted
has little effect on the simulation results, which change by less than 0.2\%
for 0-5\% centrality if the $R_{AA}$ measurement at $p_T=2.0$ GeV/c is 
used instead.  The latter scenario is intended to model the
effect of the suppression of jets due to energy loss in the nuclear 
medium~\cite{Wang98} on the fluctuation signal. The initial value of
$S_{prob}(N_{part})$ for both scenarios is normalized so that the $F_{p_T}$
result from the $R_{AA}$-scaled simulation matches that of the data
for the 20-25\% centrality class. The results of the simulation
as a function of $p_{T}^{max}$, with $S_{prob}(N_{part})$ scaled by
$R_{AA}$, are represented by the dashed curve in Fig. 3 for the 20-25\% 
centrality class,  The trend of increasing $F_{p_T}$ with
increasing $p_{T}^{max}$ observed in the data is reproduced by the
simulation reasonably well.

The results of the two hard scattering simulation scenarios are shown 
in Fig. 2 as a function of centrality.  The model curves include the small
contribution estimated from the elliptic flow simulation.
The dotted curve is the result with
$S_{prob}(N_{part})$ fixed for all centralities. The dashed curve is the
result with $S_{prob}(N_{part})$ scaled by $R_{AA}$ as a function of
centrality.  Within this simulation, the decrease of $F_{p_T}$ for the
more peripheral events is explained as a decrease in the signal strength
relative to number fluctuations from the small and decreasing value 
of $N_{tracks}$.  If $S_{prob}(N_{part})$ remains constant,
the value of $F_{p_T}$ decreases only slightly when going from mid-central to
central collisions, in contradiction with the large decrease seen in the data
over this centrality range.  When $S_{prob}(N_{part})$ is scaled by
$R_{AA}$ as a function of centrality, the trend in the simulation of
decreasing $F_{p_T}$ with increasing centrality is more consistent with the
data.

To summarize, the PHENIX experiment has observed a positive non-random
fluctuation signal in event-by-event average transverse momentum, measured as a
function of centrality and $p_T$ in $\sqrt{s_{NN}}=200$ GeV Au+Au and
p+p collisions.  The increase of $F_{p_T}$ with increasing $p_T$ 
implies that the majority of the fluctuations are due to correlated high 
$p_T$ particles.  
A Monte Carlo simulation that includes elliptic flow and a PYTHIA-based
hard scattering description can consistently describe contributions to the 
signal as a function of centrality and $p_T$ with a simple implementation of 
jet suppression.

%\section{Acknowledgements}

%\section{Acknowledgements}   % Run-2 short form for PRL

We thank the staff of the Collider-Accelerator and Physics
Departments at BNL for their vital contributions.  We acknowledge
support from the Department of Energy and NSF (U.S.A.), MEXT and
JSPS (Japan), CNPq and FAPESP (Brazil), NSFC (China), CNRS-IN2P3
and CEA (France), BMBF, DAAD, and AvH (Germany), OTKA (Hungary),
DAE and DST (India), ISF (Israel), KRF and CHEP (Korea),
RMIST, RAS, and RMAE, (Russia), VR and KAW (Sweden), U.S. CRDF
for the FSU, US-Hungarian NSF-OTKA-MTA, and US-Israel BSF.

% Reference list must begin with these first four lines:

%FIGURES:  Place all the figures here (after the references) in sequence.

\end{document}